\documentclass[prl,twocolumn,preprintnumbers,amsmath,amssymb,superscriptaddress]{revtex4}

\usepackage{graphicx}
\usepackage{dcolumn}
\usepackage{bm}
\usepackage{braket}
\usepackage{url}

\begin{document}

\preprint{}

\title{Dirac and topological phonons with spin-orbital entangled orders}
\author{Yan-Qi Wang}
\affiliation{International Center for Quantum Materials, School of Physics, Peking University, Beijing 100871, China}
\affiliation{Collaborative Innovation Center of Quantum Matter, Beijing 100871, China}
\author{Xiong-Jun Liu}
\email{xiongjunliu@pku.edu.cn}
\affiliation{International Center for Quantum Materials, School of Physics, Peking University, Beijing 100871, China}
\affiliation{Collaborative Innovation Center of Quantum Matter, Beijing 100871, China}

\date{\today}

\begin{abstract}
We propose to study novel quantum phases and excitations for a 2D spin-orbit (SO) coupled bosonic $p$-orbital optical lattice based on the recent experiments. The orbital and spin degrees of freedom with SO coupling compete and bring about nontrivial interacting quantum effects. We develop a self-consistent method for bosons and predict a spin-orbital entangled order for the ground phase, in sharp contrast to spinless high-orbital systems. Furthermore, we investigate the Bogoliubov excitations, showing that the Dirac and topological phonons are obtained corresponding to the predicted different spin-orbital orders. In particular, the topological phonons exhibit a bulk gap which can be several times larger than the single-particle gap of $p$-bands, reflecting the enhancement of topological effect by interaction. Our results highlight the rich physics predicted in SO coupled high-orbital systems and shall attract experimental efforts in the future.
\end{abstract}

\maketitle

\emph{Introduction--}Ultracold atom gases offer unique platforms for exploring many-body physics due to their fully controllable manner~\cite{Bloch2012}. The recently developed technologies allow the exploitation of orbital and spin, as two fundamental degrees of freedom of an atom, to simulate exotic quantum phases. In particular, the spin-orbit (SO) coupling for ultracold atoms can be engineered by Raman techniques which induce spin-flip transitions and momentum transfer simultaneously~\cite{Osterloh2005,Ruseckas2005,Liu2009prlEffect,LXJPRL2014,Baozong2017}. The experimental realizations of one-dimensional (1D)~\cite{Lin2011NatureSOCBEC,Zhang2012prlCollective,Wang2012prlSO,Cheuk2012prlSpininjection,Qu2013praOservation,SongBoPhysRevA2016,Lev2016PhysRevX} and 2D~\cite{huang2016experimental,Wu2016science2D,meng2016experimental,Sunwei2017} SO couplings for ultracold atoms open up extensive studies of novel quantum physics, including spintronic effects~\cite{Jairo2015RevModPhys}, magnetic and stripe phases~\cite{WuCPL2011stripe,Zhai2010PRLstripe,HoPRL2011stripe}, topological insulators and topological superfluids~\cite{Hasan2010,QiZhang2011,Zhang2008,Sato2009,Goldman2014review,Yi2015,Goldman2016,Chan2017PhysRevLetta,Chan2017PhysRevLett}, which have attracted considerable attention in ultracold atom physics. With the advantages of detection and manipulation, both the equilibrium and non-equilibrium quantum phases of nontrivial topology have been reported in ultracold atom experiments~\cite{Goldman2015Nature,Floquet2016,JoGyuBoong2017}.

In optical lattice SO coupled ultracold atoms have been considered in $s$-orbital regime. On the other hand, the high-orbital states, e.g. the $p$-orbitals, exhibit nontrivial orbital degree of freedom for having an intrinsic orbital degeneracy in optical lattices~\cite{Isacsson2005pra}. More than simulating complex physics of electrons in crystals, quantum phases for ultracold orbital bosons have no prior analogue in solid state systems~\cite{Li2016RPPreview}. The early theory predicted that interacting $p$-orbital bosons can spontaneously break time-reversal symmetry and form local orbital angular-momentum order~\cite{LiuWu2006pra,Kuklov2006prl,Wu2006prlQuantum,Wu2009MPLBUnconventional}, which further renders chiral bosonic Mott phases in strong interacting regime~\cite{LarsonPhysRevA2009,Li2012PhysRevLett,LewensteinPhysRevLett2013}. The $p$-orbital states with nontrivial topology~\cite{Wu2008prlOrbital,LiuPhysRevA2010,Xu2016PhysRevLett,LibertoPhysRevLett2016} and novel orders for fermions~\cite{ZhaoPRL2008,CaiPhysRevA2011} were also proposed. Importantly, some of the bosonic orbital phases have been successfully observed in the recent experiments~\cite{Wirth2011NatPhy,Olschlager2013NJP,Kock2015prl}.

Motivated by the experimental progresses of both SO coupling and high-orbital physics for ultracold atoms, we propose in this letter to investigate the emergence of novel interacting phases for a $p$-orbital SO coupled bosonic optical lattice. Compared with a spinless $p$-orbital system, the inclusion of SO coupling introduces spin degree of freedom competing with the orbital one and brings about fundamentally new quantum physics. In particular, we predict a novel {\it spin-orbital entangled} ordering for the ground phase, and further show that the Dirac and topological Bogoliubov phonons can be obtained corresponding to the different spin-orbital orders. These results are observable based on the current experiments.

\begin{figure}[!t]
\centering
\includegraphics[width=0.9\columnwidth]{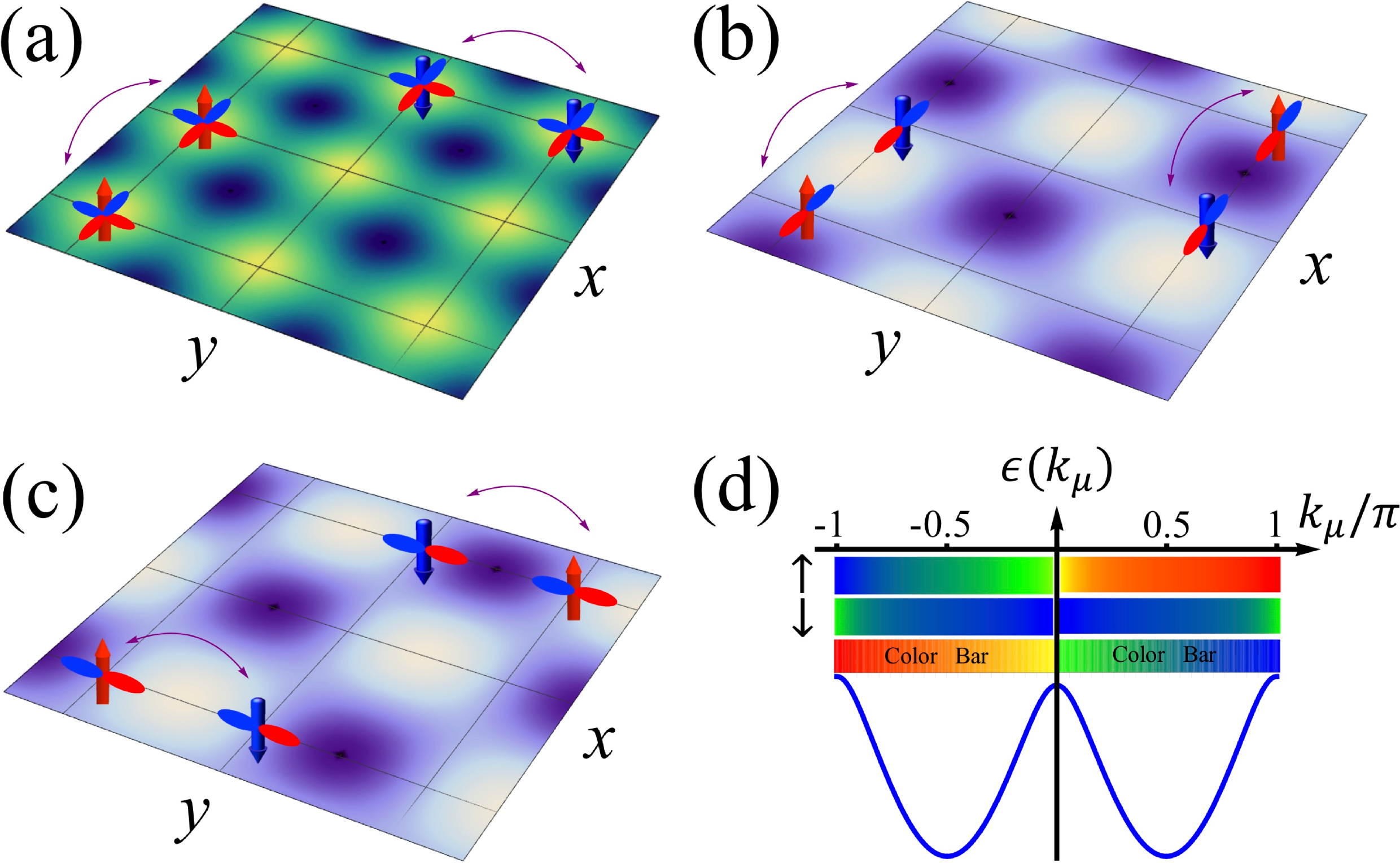}
\caption{\label{socandporbit}(Color online) Spin-1/2 Bosonic $p$-orbital optical lattice, with vertical red and blue arrows denotes spin-up and spin-down, respectively. (a) Sketch of spin-conserved neighboring hopping for $p$-orbitals induced by lattice potential; (b) Raman potential $M_x$ and the induced spin-flipped hopping along $x$ direction; (c) Raman poential $M_y$ and the induced spin-flipped hopping along y direction; (d) The energy spectrum for the non-interacting Hamiltonian $H_{0\mu}$, with $t_{so} = 3t_0$ and $m_z = 0.1t_0$. The colored stripes show the spin-polarization of Bloch states in the first Brillouin zone.}
\end{figure}

\emph{The model}.--We start with the $p$-orbital ultracold bonsons trapped in a 2D square optical Raman lattice which was realized in recent experiments~\cite{Wu2016science2D,Sunwei2017}, with the lattice potential $V_{\text{latt}}(x,y) = V_0 (\cos^2 k_0x + \cos^2 k_0 y)$ and the periodic Raman potential $V_R(x,y) = M_0 \cos k_0 x \sin k_0 y \sigma_x+M_0 \cos k_0 y \sin k_0 x \sigma_y$, as illustrated in Fig.~\ref{socandporbit}. Here $\sigma_{x,y}$ are Pauli matrices defined in atomic spin space, $V_0, M_0$, and $k_0$ are lattice depth, Raman coupling amplitude, and wave vector of Raman beams, respectively. The total Hamiltonian $H = H_0 + H_{{I}}$, where $H_0$ ($H_I$) denotes the single-particle (interacting) term, as given below. In the tight-binding regime, the $p_{\mu}$ ($\mu=x,y$) orbital has a dominating hopping along $\mu$ direction ($\sigma$-bonding), while the hopping along traverse direction ($\pi$-bonding) is minimized (Fig. \ref{socandporbit}). We keep only $\sigma$-bonding for the tight-binding limit, in which case $H_0=\sum_\mu H_{0\mu}$, with
\begin{eqnarray}\label{eqn:tightbindingSI1}
H_{0\mu} &=&-\sum_{\langle\vec{i},\vec{j}\rangle_\mu,s,s'}t_{\mu}\hat p_{\vec{i}s \mu}^{\dag}\sigma_z^{ss'}\hat
p_{\vec{j}s' \mu}+\sum_{\vec{i}}m_z(\hat n_{\vec{i}\uparrow\mu }-\hat n_{\vec{i}\downarrow\mu }) \nonumber \\
&&+\bigr[\sum_{\langle\vec i,\vec j\rangle_\mu}\delta_\mu t_{\rm so}(\hat p_{\vec i\uparrow\mu }^\dag\hat p_{\vec j \downarrow \mu }-\hat p_{\vec i\uparrow \mu }^\dag\hat p_{\vec j \downarrow\mu})+{\rm H.c.} \bigr].
\end{eqnarray}
Here $p_{\vec i s \mu}^\dagger$ and $p_{\vec i s \mu }$ are creation and annihilation operators for a boson at lattice site $\vec i$, with spin $s=\uparrow,\downarrow$ and orbital $p_\mu$; $\langle \vec i, \vec j\rangle_\mu$ stands for nearest-neighbour hopping along $\mu$ direction; $t_\mu$ and $t_{\text{so}}$ denotes spin-conserved and spin-flipped, respectively; $m_z$ is an effective Zeeman term, $\delta_x = 1$ and $\delta_y=i$. Transforming $H_0$ into momentum space yields $ H_0 = \sum_{\bm{k} s s^\prime \mu}\hat p^\dagger_{\bm{k}s \mu} \mathcal {H}_{0\mu}^{ss'}(\bm{k}) \hat p_{\bm{k}s^\prime \mu}$, with
\begin{equation}\label{singleorigin}
\mathcal H_{0\mu}(\bm{k}) = [m_z + 2t_\mu \cos (k_\mu a)]\sigma_z - 2t_{\text{so}} \sin (k_{\mu} a) \sigma_\mu,
\end{equation}
where $a$ is the lattice constant. For the isotropic lattice potential, we have $t_x = t_y =- t_0$. For our purpose we consider in this work the large SO coupling regime with $t_{\rm so}^2 > t_0^2 + t_0|m_z|/2 $, so that the lower subband of the Bloch Hamiltonian $\mathcal H_{0\mu}(\bm{k})$ has two minima in first Brillouin zone. The band structure and spin polarization are shown in Fig. [\ref{socandporbit}].

The interaction for our consideration is short-range repulsive, with the interacting coefficients $g_{\uparrow\uparrow}=g_{\uparrow\downarrow}=g_{\downarrow\downarrow}=g$ between bosons. Thus interacting part of the Hamiltonian is shown to take the following form~\cite{supp}
\begin{equation}\label{intorigin}
H_{{I}} = U_{\text{int}} \sum_{\vec {i}} \{ n^2_{\vec {i}} +\frac{1}{2} \sum_{\mu s} (n^2_{\vec i, \mu s} + p^\dagger_{\vec i, \mu s}p^\dagger_{\vec i,\mu s} p_{\vec i,\bar \mu s} p_{\vec i,\bar \mu s}) \},
\end{equation}
with $U_{\text{int}} =g \int d^3 \vec r |\phi_{p_\mu}(\vec r)|^4 >0$, $n_{i,\mu s}=p^\dagger_{\vec i, \mu s}p_{\vec i,\mu s}$, and $\mu\neq\bar \mu$. Without the SO Hamiltonian $H_0$, the ground state of interacting Hamiltonian $H_I$ solely is given as follow: the orbital part of both spin-up and spin-down atoms is an eigenstate of the local angular momentum $L_z = -i(p_x^\dagger p_y - p^\dagger_y p_x)$, i.e. having the local orbital configuration $\frac{1}{\sqrt{2}}(p_x + ip_y)_s$ or $\frac{1}{\sqrt{2}}(p_x-ip_y)_s$, similar to the results obtained for spinless bosons~\cite{LiuWu2006pra,Kuklov2006prl,Wu2006prlQuantum,Wu2009MPLBUnconventional,You2016PhysRevA}. As uncovered below, further inclusion of SO coupling brings about nontrivial interplay between spin and $p$-orbital degrees of freedom, leading to new interacting quantum phases.

\emph{Ground phase--}We now turn to solving the ground state of the total Hamiltonian with both non-interacting part $H_0$ and interacting part $H_I$. With both spin and orbital degrees of freedom, the direct calculation of the ground phase of the present system is not convenient. We develop a self-consistent method which is useful for the system of multi-component bosons. The self-consistent order parameters are introduced as $\Delta_\uparrow = \langle p^\dagger_{x(y)\uparrow}p_{x(y)\uparrow}\rangle_G,~\Delta_{\downarrow}=\langle p^\dagger_{x(y)\downarrow} p_{x(y)\downarrow} \rangle_G, ~ \Delta_{xy\uparrow} = \langle p^\dagger_{x\uparrow} p_{y\uparrow}  \rangle_G,~\Delta_{xy\downarrow} = \langle p^\dagger_{x\downarrow} p_{y\downarrow} \rangle_G$, where $\langle\cdots\rangle_G$ denotes the calculation on the ground state. It can be seen that $\Delta_{\uparrow(\downarrow)}$ relate to the magnetization, and the orders $\Delta_{xy\uparrow(\downarrow)}$ stand for the hybridization between $p_x$ and $p_y$ orbitals. Note that in the current stage we do not consider the Bogoliubov quasiparticles, which will be studied later for excitations. Since our lattice has $C_4$ symmetry, we can write down the ground state as $\frac{e^{i\xi}}{\sqrt{2}}(p^\dagger_x + e^{i\varphi}p_y^\dagger)_\uparrow$ for spin-up component, and $\frac{1}{\sqrt{2}} (p^\dagger_x + e^{i\theta}p_y^\dagger)_\downarrow$ for spin-down component, where $\varphi$ and $\theta$ are the phases difference between $p_x$ and $p_y$ orbitals, with $-\pi/2 \leqslant \varphi,\theta \leqslant \pi/2$, and $\xi$ is the relative phase between spin-up and spin-down components. A direct analysis reveals that the non-interacting part $H_0$ of the Hamiltnoian favors the phase configuration with $\xi=0 $ and $\varphi-\theta=\pi/2$, while the interacting term $H_{I}$ favors $\varphi = \pm \pi/2 ,\theta = \pm \pi/2$, and arbitrary $\xi$ due to the $SO(2)$ symmetry. This implies the competition between SO term and orbital degree of freedom with interactions.

\begin{figure}[!h]
\centering
\includegraphics[width=1\columnwidth]{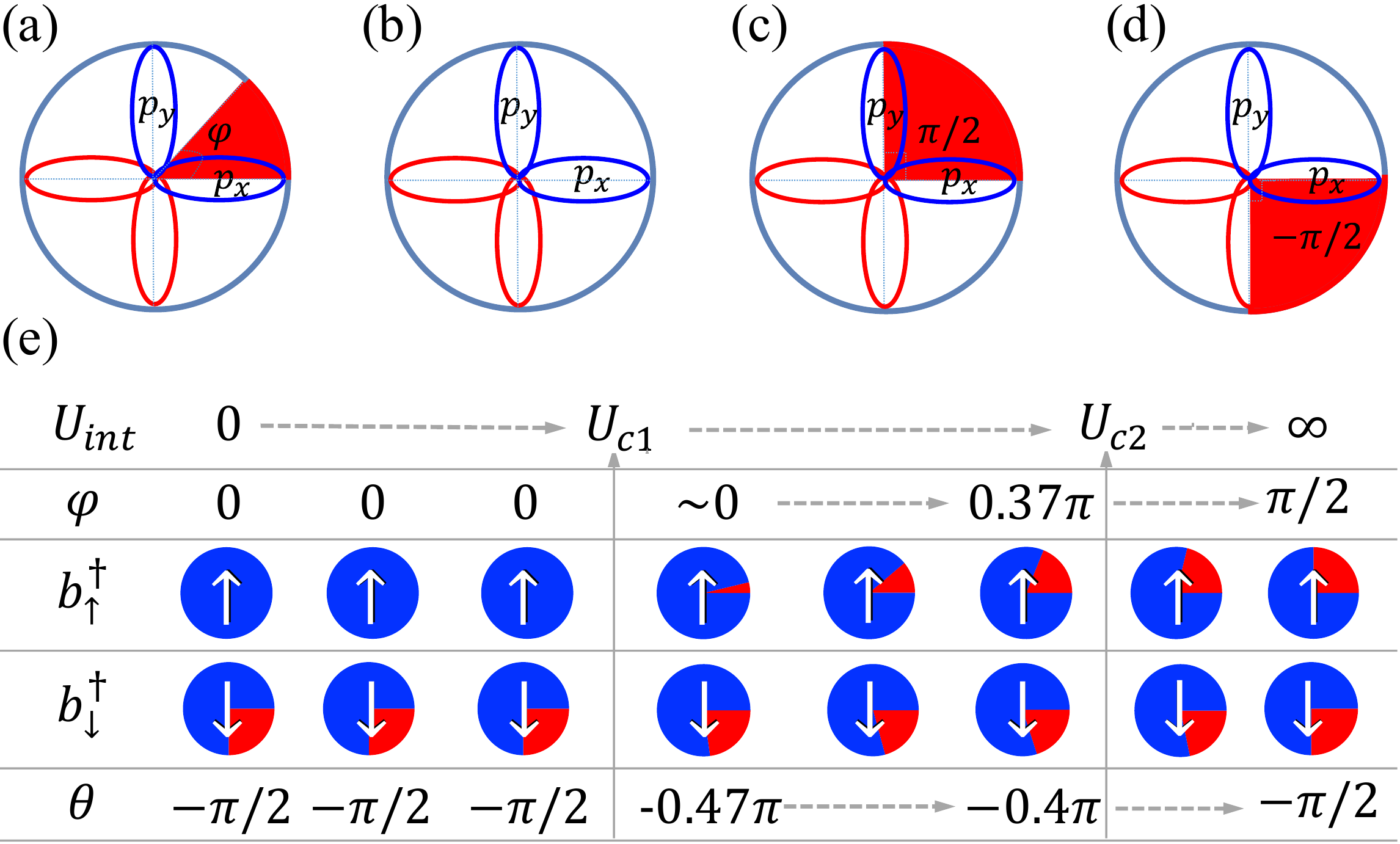}
\caption{\label{phasediagram}(Color online) Local orbital configuration for (a) $\frac{1}{\sqrt{2}} (p_x^\dagger + e^{i\varphi} p_y^\dagger)$, (b) $\frac{1}{\sqrt{2}} (p_x^\dagger +  p_y^\dagger)$, (c) $\frac{1}{\sqrt{2}} (p_x^\dagger +  i p_y^\dagger)$, and (d) $\frac{1}{\sqrt{2}} (p_x^\dagger -i   p_y^\dagger)$. The red azimuthal angle denotes the phase difference between $p_x$ and $p_y$ orbitals, with (a) the generic local orbital configuration and (c-d) the eigenstates of local angular momentum $L_z$. (e) Formation of orbital orders versus the strength of onsite interaction, with $m_z = 0.1 t_0$ and $t_{\text{so}} = 3 t_0$.}
\end{figure}

The self-consistent solution is obtained by the following iteration. First, we obtain the initial ground state $|G_{0}\rangle$ the single-particle Hamiltonian $H_0$, which corresponds to one of the band minimums of the $p$-bands. Furthermore, we compute the mean-field order parameters based on $|G_{0}\rangle$, and substitute them into total Hamiltonian to obtain mean-field Hamiltonian $H_{\rm MF}$ by linearizing the interacting term $H_I$. Then we diagonalize $H_{\rm MF}$ and obtain the new mean-field ground state $|G_{\rm MF}\rangle$, with which we recompute the order parameters, and repeat the above steps until the solved order parameters converge~\cite{supp}. Three regions of the phases are obtained, as shown numerically in Fig.~\ref{phasediagram} with the parameters $t_{\text{so}} = 3t_0$ and $m_z = 0.1t_0$. First, for the weak interacting regime with $U_{\rm int}$ is below a critical value $U_{c1}$, the orbital ordering is dominated by the SO coupling, with the phase factors $\xi = 0, \varphi =0, \theta = -{\pi}/{2}$. This follows that the spin-up component forms a purely real orbital $\frac{1}{\sqrt{2}}(p_x^\dagger + p_y^\dagger)_\uparrow$ which breaks $SO(2)$ symmetry, while the spin-down component forms an angular-momentum order $\frac{1}{\sqrt{2}}(p_x^\dagger-ip_y^\dagger)_\downarrow$, which preserves the $SO(2)$ symmetry. Secondly, when the interaction is greater than $U_{c1}$, one has $\xi \neq 0, 0<\varphi< \pi/2, -\pi/2<\theta <0$, and the orbitals of the spin-up and spin-down atoms are both imaginary. This is the moderate region, which reconciles the minimization of both non-interacting and interacting mean field energies. Both spin-up and spin-down components breaks the $SO(2)$ symmetry, indicating a phase transition at $U_{\rm int}=U_{c1}$. Beyond this value, the phase $\varphi$ for spin-up component gradually increases to $\pi/2$ with increasing interaction, while the phase $\theta$ for the spin-down component starts to deviate from $-\pi/2$ to a certain value $-\pi/2<\theta <0$. Finally, when interaction is even greater than a second quasi-critical value $U_{c2}$, the phase $\theta$ approaches $-\pi/2$ again with increasing interaction. This is a crossover region dominated by both SO coupling and interactions. The spin-up and spin-down components form $p_x+ip_y$ and $p_x-ip_y$ orders, respectively in the strong interacting limit, i.e. $\varphi = \pi/2$ and $\theta = - \pi/2$, restoring the $SO(2)$ symmetry.

The wave function of the ground state can be obtained from the self-consistently solved mean-field Hamiltonian, which is projected onto the orbital bases $b^\dagger_\uparrow=\frac{1}{\sqrt{2}}(p_x^\dagger+e^{i\varphi }p_y^\dagger)_\uparrow$ and $b^\dagger_\downarrow=\frac{1}{\sqrt{2}}(p_x^\dagger+e^{i\theta }p_y^\dagger)_\downarrow$ for spin-up and spin-down states, respectively. Writing down in $\bold k$ space we obtain $H_{\rm MF} = \sum_{\bm{k}ss^\prime} b^\dagger_{\bm{k}s} \hat h_{\rm MF}(\bm{k}) b_{\bm{k}s^\prime}$, with $\hat h_{\rm MF} (\bm k)= (\tilde m_z + t_0 \cos k_x + t_0 \cos k_y)\sigma_z -t_{\rm so}[ \sin k_x + \sin (\varphi -\theta)\sin k_y]\sigma_x-t_{\rm so}\cos(\varphi - \theta)\sin k_y \sigma_y$,
where the 2D SO coupling term and $\tilde m_z=m_z+\langle\cos^2\varphi n_\uparrow-\cos^2\theta n_\downarrow\rangle_G/2$ are corrected by mean-field quantities $\Delta_{\uparrow,\downarrow}$, $\varphi$, and $\theta$. For the large SO coupling regime the lower band has four band minimums at $\{\bold\Lambda_j\}=\{(\pm\pi/2,\pm\pi/2)\}$. The ground state corresponds to that the bosons condense at one of the four minimums, say at $\bold\Lambda_1=(\pi/2,\pi/2)$, whose single-particle state is given by $|u_{\rm min}(\bold \Lambda_1)\rangle=[\sin\alpha e^{i\beta} b^\dagger_\downarrow-\cos\alpha b^\dagger_\uparrow]|{\rm vac}\rangle$, with $\tan(2\alpha)=\tilde m_z/\sqrt{2t^2_{\rm so}[1+\sin(\varphi-\theta)]}$ and $\tan\beta=\cos(\varphi-\theta)/[1+\sin(\varphi-\theta)]$. The condensate wave function then reads
\begin{eqnarray}\label{condensate1}
|\Phi_{\rm BEC}\rangle=\sin\alpha e^{i\beta}|\Phi_{p_x+e^{i\varphi}p_y},\uparrow\rangle-\cos\alpha|\Phi_{p_x+e^{i\theta}p_y},\downarrow\rangle,
\end{eqnarray}
which is generically an entangled order between orbital and spin states since $\theta$ and $\varphi$, as shown in Fig.~\ref{phasediagram}, are different. If extrapolating to strong interacting regime, the above condensate wave function renders a maximally entangled phase: $(|\Phi_{p_x+i p_y}\rangle\otimes|\uparrow\rangle-e^{i\pi/4}|\Phi_{p_x-ip_y}\rangle\otimes|\downarrow\rangle)/\sqrt{2}$. The spin-orbital entangled phase shown in Eq.~\eqref{condensate1} is a main result predicted in this letter, uncovering the new exotic states resulted from the interplay of SO coupling and orbital degree of freedom with interactions.

\emph{Bogoliubov excitations--}Having obtained the ground state of the $p$-orbital condensate, we can further investigate the physics of the Bogoliubov quasiparticles, i.e. the phonons. Note that one cannot study the excitations by the effective Hamiltonian $\hat h_{\rm MF}(\bm k)$, which is used to self-consistently solve the ground state only. On the other hand, since the ground phase only involves the two modes $b_\downarrow$ and $b_\uparrow$, in studying the phonons we project the interacting Hamiltonian onto the bases $b_{\uparrow,\downarrow}$ and reach the following effective form
\begin{equation}
H_I^\prime = U_{\text{int}}\sum_{\vec i} (n^2_{\vec i}+ \frac{1}{2}n^2_{\vec i,\uparrow} \cos^2 \varphi+\frac{1}{2}n^2_{\vec i,\downarrow}\cos^2\theta),
\end{equation}
where $n_{\vec i s} = b^\dagger_{\vec i s}b_{\vec i s}$ and $n_{\vec i}=n_{\vec i \uparrow}+n_{\vec i \uparrow}$. From the above formula one can find that under the bases $b_{\uparrow,\downarrow}$ the effective interaction coefficients between bosons becomes $g_{\uparrow\uparrow} =g(1 + \frac{1}{2}\cos^2 \varphi)$, $g_{\downarrow\downarrow}=g(1 + \frac{1}{2}\cos^2\theta)$ and $g_{\uparrow\downarrow}=g$.

We solve the Bogoliubov excitations by taking that the bosons are condensed in the state of Eq.~\eqref{condensate1}. Following the standard Bogoliubov theory, the field operator can be expanded as $\hat \psi_s = \phi_s + \delta \hat \psi_s$, where $\phi_s=\langle b_{\vec i s}\rangle_G$ is the spin-$s$ component of the ground state wavefunction and $\delta \hat \psi_s$ denotes the fluctuation. The Bogoliubov Hamiltonian under the basis $(\delta \hat \psi_\uparrow , \delta \hat \psi_\downarrow, \delta \hat \psi_\uparrow^\dag,\delta\hat \psi_\downarrow^\dag)$ is given by~\cite{PL2016,Pan2016}
\begin{equation}\label{bog1}
{\mathcal {\hat {H}}_B}=
\begin{pmatrix}
\hat h_0(\bm{k}) + \Gamma_1 - \mu & \Gamma_2 \\
\Gamma^*_2 & \hat h_0^*(-\bm{k}) + \Gamma_1^* -\mu
\end{pmatrix}
\end{equation}
with:
\begin{equation}
\Gamma_1=
\begin{pmatrix}
2g_{\uparrow\uparrow} |\phi_{\uparrow}|^2+g_{\uparrow\downarrow}|\phi_\downarrow|^2 & g_{\uparrow\downarrow}\phi_\uparrow\phi_\downarrow^* \\
g_{\uparrow\downarrow}\phi_{\downarrow}\phi_{\uparrow}^* & g_{\uparrow\downarrow}|\phi_{\uparrow}|^2+2g_{\downarrow\downarrow}|\phi_\downarrow|^2
\end{pmatrix},\nonumber
\end{equation}
\begin{equation}
\Gamma_2=
\begin{pmatrix}
g_{\uparrow\uparrow}\phi_{\uparrow}^2 & g_{\uparrow\downarrow}\phi_{\uparrow}\phi_{\downarrow} \\
g_{\uparrow\downarrow}\phi_{\downarrow}\phi_{\uparrow} & g_{\downarrow\downarrow} \phi_{\downarrow}^2
\end{pmatrix}.\nonumber
\end{equation}
The chemical potential $\mu$ in Eq.~\eqref{bog1} can be determined numerically through imaginary-time evolution~\cite{supp}. Then one can solve the spectra and eigenstates of excitations by diagonalizing the Bogoliubov Hamiltonian ${\mathcal {\hat {H}}_B}$ with a para-unitary transformation $\hat T_{\bold k}$ for bosons~\cite{Shindou1,Shindou2,Ueda2015}:
\begin{equation}
\hat \sigma_3 {\mathcal {\hat {H}}_B} \hat T_{\bold k} = \hat T_{\bold k} \hat \sigma_3 E_{d,{\bold k}}, \quad \sigma_3 = \text{diag} \{ 1, 1, -1, -1 \}.
\end{equation}
The topology of the $n$-th phonon band can be determined
by the Chern number from integral of Berry curvature
\begin{eqnarray}\label{BerryCurvatureModify}
C_{1}^{(n)} &=& -\frac{1}{2\pi} \int_{\text{FBZ}} dk_x dk_y \Omega^{xy}_{n,{\bold k}},\\
\Omega^{xy}_{n,k}&=&i(\sigma_3)_{n,n}\epsilon_{xy}(\frac{\partial}{\partial_{k_x}} \bra{t_{n}(\bm k)})\sigma_3 (\frac{\partial}{\partial_{k_y}} \ket{t_n(\bm k)}),\nonumber
\end{eqnarray}
where $\ket{t_n(\bm k)}$ is the $n$-th column vector of $\hat T_{\bold k}$. All the results can be solved numerically.
\begin{figure}[!h]\centering
\includegraphics[width=0.95\columnwidth]{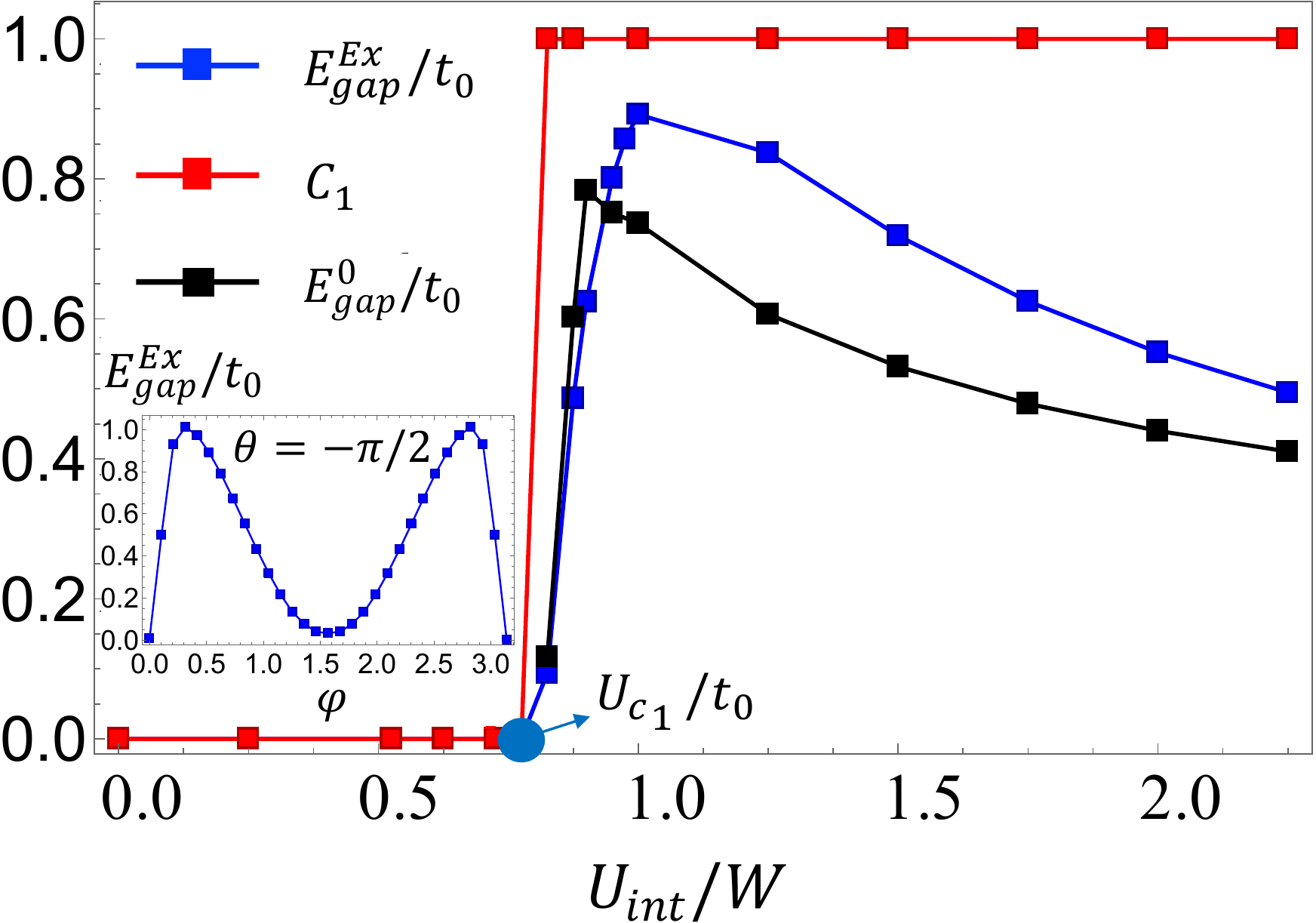}
\caption{\label{gap}(Color online) The blue and black lines denote gaps for phonon excitation and the Hamiltonian $\hat h_{\rm MF} (\bm k)$, respectively, with $m_z = 0.1t_0$, $t_{\text{so}} =3t_0$, and $W=4t_0$. The red line denotes the band topology of phonons. The inserted subfigure shows the phonon gap versus $\varphi \neq 0$ with fixed $\theta = -\pi/2$.}
\end{figure}

\begin{figure}[!h]\centering
\includegraphics[width=0.95\columnwidth]{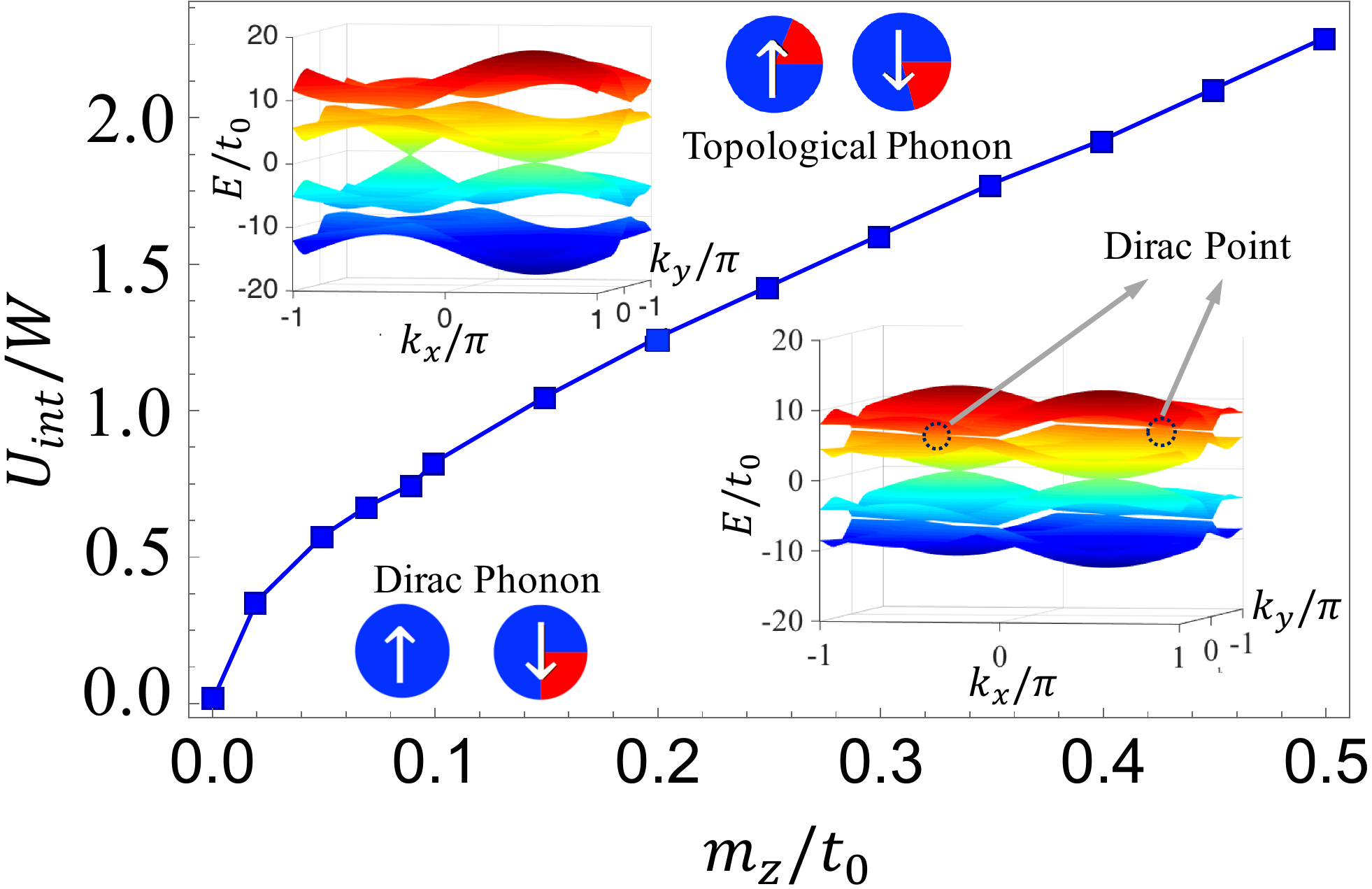}
\caption{\label{umz}(Color online) The phase diagram plotted versus interaction $U_{\rm int}$ and $m_z$, with $t_{\text{so}}=3t_0$. The blue curve gives the critical interaction $U_{c1}$. The Dirac (topological) phonon spectra correspond to the parameter regime $m_z=0.2 t_0$, $t_{so} = 3t_0$ and $U<U_{c1}$ ($m_z = 0.2$, $t_{so} =3t_0$ and $U>U_{c1}$).}
\end{figure}

Fig.~\ref{gap} shows the numerical results for the phonon band gap ($E_{\rm gap}^{\rm ex}$, blue line) and the Chern number of the lowest band ($C_1$, red line). For comparison, we also plot the band gap $E_{\rm gap}^0$ corresponding to the Hamiltonian $\hat h_{\rm MF} (\bm k)$ (black line). When $U_{\text{int}} < U_{c1}$, the excitation spectra are gapless, exhibiting two anisotropic Dirac cones for the Dirac phonons (see also Fig.~\ref{umz}). According to Eq.~\eqref{condensate1}, in this regime the spin-orbital entangled ground state is a superposition of $|\Phi_{p_x+ p_y}\rangle\otimes|\uparrow\rangle$ and $|\Phi_{p_x-ip_y}\rangle\otimes|\downarrow\rangle$. When $U_{\text{int}} > U_{c1}$, the orbitals of both spin-up and spin-down components are imaginary, with the excitation bulk gap $E_{\rm gap}^{\rm ex}$ and single-particle gap $E_{\rm gap}^0$ opening up. The Chern number $C_1$ is nonzero, implying that the phonon band is topologically nontrivial. Interestingly, the bulk gap $E_{\rm gap}^{\rm ex}$ of phonons (also $E_{\rm gap}^0\sim2\tilde m_z$) increases quickly with $U_{\rm int}$ and exhibits a maximum value which is several times larger than the single-particle gap ($2m_z=0.2t_0$) of $p$-bands, showing the considerable enhancement of the topological gap of phonons by the interactions.
The non-monotonous behavior of the phonon gap is a consequence of the modification to $(\varphi,\theta)$ and thus $\tilde m_z$ by the interactions. As we plot the phonon bulk gap versus $\varphi$ with fixed $\theta = -\pi/2$ (inserted picture in Fig.~\ref{gap}), a maximum of the gap is clearly obtained with $\varphi\approx0.1\pi$.

The whole phase diagram is further shown in Fig.~\ref{umz} versus $m_z$ and $U_{\rm int}$. The blue line denotes the critical value of phase transition ($U_{c1}$) with different $m_z$. Below the transition line the phonons are of Dirac type with two gapless Dirac points in the bulk, and above the blue line is the region for topological phonons with a bulk gap (for $m_z>0$). One can see that $U_{c1}$ increases as $|m_z|$ increases. A special case occurs at zero magnetization $m_z = 0$, where the system is symmetric with respect to the exchange of spin-up and spin-down. For this one can show straightforwardly that $U_{c1} = U_{c2} = 0$, and the spin-up and spin-down components would be pined to $p_x+e^{i\phi}p_y$ and $p_x+e^{-i\phi}p_y$ orbitals for finite $U_{\rm int}$. The critical values $U_{c1,c2}$ increase with increasing magnetization.

\emph{Conclusion--}In conclusion, we have predicted a novel spin-orbital entangled phase in a 2D SO coupled bosonic $p$-orbital optical lattice, and further shown that the Dirac and topological phonons are obtained with different spin-orbital orders. The predicted new phases are a consequence of the interplay between the SO coupling and high-orbital states with interactions, revealing the rich physics resulted by taking into account both the spin and orbital degrees of freedom for bosons. Many interesting issues deserve attention following this prediction, including SO coupled $p$-orbital bosonic Mott phases which might exhibit novel physics beyond the current knowledge and SO coupled $p$-orbital Fermi systems. With the recent experimental progresses of realizing 2D SO coupling~\cite{Wu2016science2D} with experimentally confirmed long lifetime~\cite{Baozong2017,Sunwei2017} and $p$-orbital superfluids~\cite{Wirth2011NatPhy,Olschlager2013NJP,Kock2015prl} in optical lattices, this work may motivate broad studies of orbital physics with SO coupling and in particular, shall attract the experimental efforts in the future.

\emph{Acknowledgement--} We appreciate the valuable discussions of Congjun Wu, Jian-Song Pan, and Long Zhang. This work is support by MOST (Grant No. 2016YFA0301604), NSFC (No. 11574008), and Thousand-Young-Talent Program of China.

\bibliography{reference}

\onecolumngrid

\renewcommand{\thesection}{S-\arabic{section}}
\setcounter{section}{0}  
\renewcommand{\theequation}{S\arabic{equation}}
\setcounter{equation}{0}  
\renewcommand{\thefigure}{S\arabic{figure}}
\setcounter{figure}{0}  

\indent

\begin{center}\large
\textbf{Supplementary Material:\\ Dirac and topological phonons with spin-orbital entangled orders}
\end{center}

In this Supplementary Material we provide more details for deriving the interacting Hamiltonian, self-consistent solution, and topological excitations.

\section{Interacting Hamiltonian}

We derive the interacting Hamiltonian for spin-$1/2$ $p$-orbitals with repulsive short range interaction $g\delta(\bf{r_1} - \bf{r_2})$. It is convenient to use the local angular-momentum bases for the the calculation. The creation and annihilation operators in the angular momentum bases take the forms
\begin{equation}
u_\uparrow^\dagger = \frac{1}{\sqrt 2}(p_x^\dagger + ip_y^\dagger)_\uparrow, \quad u_\downarrow^\dagger = \frac{1}{\sqrt 2}(p_x^\dagger + ip_y^\dagger)_\downarrow, \quad v^\dagger_\uparrow = \frac{1}{\sqrt 2}(p_x^\dagger - ip_y^\dagger)_\uparrow, \quad v^\dagger_\downarrow =\frac{1}{\sqrt 2} (p_x^\dagger - ip_y^\dagger)_\downarrow,
\end{equation}
which obey the commutation relations of bosons $[u_\sigma, u^\dagger_{\sigma^\prime}] =
\delta_{\sigma \sigma^\prime}$ and $[v_\sigma, v^\dagger_{\sigma^\prime}] = \delta_{\sigma \sigma^\prime}$. Note that the short range repulsive contact interaction $g\delta(\bf {r_1} - \bf{r_2})$ conserves the spin and local angular momentum. The effective interacting Hamiltonian includes the following scattering processes.

First, the two bosons of interaction have total local angular momentum $\langle L_z \rangle = 2~(-2)$, which corresponds to the process $\ket{uu} \rightarrow \ket{uu}$ ($\ket{vv} \rightarrow \ket{vv}$). Depending on whether the spin states of the two bosons are the identical or different, these scattering processes include the following two cases. The first is the scattering between the same spin states, i.e, $\ket{u_\sigma u_\sigma} \rightarrow \ket{u_\sigma u_\sigma}$. The interacting coefficient for this process is calculated by
\begin{eqnarray}
E_{\mbox{\tiny{int1}}} &=&  \frac{g_{\uparrow \uparrow}}{8}\int d^3 \vec r_1 d^3 \vec r_2 \bigr[\phi_{p_x}^*(\vec r_1) - i\phi_{p_y}^*(\vec r_1)\bigr]\bigr[\phi_{p_x}^*(\vec r_2) - i\phi_{p_y}^*(\vec r_2)\bigr] \delta(\vec r_1 -\vec r_2) \bigr[\phi_{p_x}(\vec r_1) + i\phi_{p_y}(\vec r_1)\bigr]\bigr[\phi_{p_x}(\vec r_2) + i\phi_{p_y}(\vec r_2)\bigr]\nonumber\\
&=& (\frac{U}{4} + \frac{V}{4}),
\end{eqnarray}
where we have denoted as $U = \int d^3 \vec r_1 |\phi_{p_x}(\vec r_1)|^4$ and $V = \int d^3 \vec r_1|\phi_{p_x}(\vec r_1)|^2 |\phi_{p_y}(\vec r_1)|^2$, with $V =U/3$~\cite{LiuWu2006pra}.
These scattering processes lead to Hamiltonian:
\begin{eqnarray}
H_{\rm int,1}^{(a)}=\sum_{\vec r,s=\uparrow,\downarrow} \frac{U+V}{4}\bigr[u^\dagger_{s}(\vec r) u^\dagger_{s}(\vec r)u_s(\vec r) u_s(\vec r) + v^\dagger_{s}(\vec r) v^\dagger_{s}(\vec r) v_s(\vec r) v_s(\vec r)\bigr].
\end{eqnarray}
The second type is the scattering between different spin-component $\ket{u_{\sigma} u_{\sigma^\prime}} \rightarrow \ket{u_\sigma u_{\sigma^\prime}},~(\sigma \neq \sigma^\prime)$. By similar calculation one can find that the scattering processes leads to Hamiltonian:
\begin{eqnarray}
H_{\rm int,1}^{(b)}=\sum_{\vec r} \frac{U+V}{2}\bigr[u^\dagger_\uparrow(\vec r) u^\dagger_\downarrow(\vec r) u_\downarrow(\vec r) u_\uparrow(\vec r) + v^\dagger_\uparrow(\vec r) v^\dagger_\downarrow(\vec r) v_\downarrow(\vec r) v_\uparrow(\vec r)\bigr].
\end{eqnarray}
With the above results, the scattering channel for the local total angular momentum $\langle L_z \rangle =\pm 2$ gives the following interacting Hamiltonian
\begin{equation}
H_{{\text{{int},1}} }=H_{\rm int,1}^{(a)}+H_{\rm int,1}^{(b)} =\frac{U+V}{4}\sum_{\vec r} \{ \bigr[n_{u\uparrow}(\vec r)+n_{u\downarrow}(\vec r)\bigr]^2 + \bigr[n_{v\uparrow}(\vec r) + n_{v\downarrow}(\vec r)\bigr]^2 \}
\end{equation}

Secondly, the two bosons of scattering have total local angular momentum $\langle L_z \rangle = 0$, which corresponds to the process $\ket{uv} \rightarrow \ket{uv}$. Similarly, we again have two different cases for the scattering between the same spin states and between opposite spin states, respectively. For the former case, we have that $\ket{u_\sigma v_\sigma} \rightarrow \ket{u_\sigma v_\sigma}$,~$(\sigma \neq \sigma^\prime)$, which leads to the interacting term
\begin{eqnarray}
H_{\rm int,2}^{(a)}=(U+V)\sum_{\vec r} \bigr[u^\dagger_\uparrow(\vec r) v^\dagger_\uparrow(\vec r) v_\uparrow(\vec r) u_\uparrow(\vec r) + u^\dagger_\downarrow(\vec r) v^\dagger_\downarrow(\vec r) v_\downarrow(\vec r) u_\downarrow(\vec r)\bigr].
\end{eqnarray}
Similarly, for the case of scattering between different spin states, namely, $\ket{u_\sigma v_{\sigma^\prime}} \rightarrow \ket{u_\sigma v_{\sigma^\prime}},~(\sigma \neq \sigma^\prime)$, the corresponding interacting term can be verified to be
\begin{eqnarray}
H_{\rm int,2}^{(b)}=\frac{U+V}{2} \sum_{\vec r}\bigr[u^\dagger_{\uparrow}(\vec r) v^\dagger_{\downarrow}(\vec r) v_{\downarrow}(\vec r) u_{\downarrow}(\vec r) + u^\dagger_{\downarrow}(\vec r)v^\dagger_{\uparrow}(\vec r)v_{\uparrow}(\vec r)u_{\downarrow}(\vec r)\bigr].
\end{eqnarray}
Then, the interacting Hamiltonian contributed by the scattering processes with local angular momentum $\langle L_z \rangle =0$ can be obtained by
\begin{eqnarray}
H_{{\text{int},2}} = (U+V) \sum_{\vec r} n_u n_v - \frac{U+V}{2}\sum_{\vec r}(n_{u\uparrow}n_{v\downarrow}+n_{u\downarrow}n_{v\uparrow})
\end{eqnarray}

Having the above results, we can now reach the total effective interacting Hamiltonian for $p$-orbital states in terms of angular-momentum bases in the following form
\begin{equation}\label{spin}
\begin{aligned}
H_{I} &= H_{\text{int1}} + H_{\text{int2}} \\
&= \frac{U+V}{4}\sum_{\vec r} (n_u^2 -n_u +n_v^2 -n_v)
+ (U+V) \sum_{\vec r} ( n_un_v -\frac{1}{2}(n_{u\uparrow}n_{v\downarrow} + n_{u\downarrow}n_{v\uparrow}) )\\
&= \frac{U+V}{4} \sum_{\vec r} (n_u^2 +n_v^2 +4n_un_v - n_u-n_v)-\frac{U+V}{2}\sum_r(n_{u\uparrow}n_{v\downarrow}+n_{u\downarrow}n_{v\uparrow})\\
&=V\sum_{\vec r} ( n^2 +2n_{u\uparrow}n_{v\uparrow}+2n_{u\downarrow}n_{v\downarrow} - n_u -n_v),
\end{aligned}
\end{equation}
where the particle number operators $n_u = n_{u\uparrow} + n_{v\downarrow}$ and $n_v = n_{v\uparrow}+n_{v\downarrow}$. Note that for the spinless system, $n_{\uparrow(\downarrow)} = n$, $n_{\downarrow(\uparrow)} =0$. One can easily reduce the Eq.~\eqref{spin} to
\begin{eqnarray}
H_{I} &=&\frac{3(U+V)}{8} \sum_{\vec r}\{ (n_u+n_v)^2  - \frac{1}{3}(n_u -n_v)^2 -\frac{2}{3}n_u-\frac{2}{3}n_v\}\nonumber\\
&=& \frac{U}{2} \sum_{\vec r} \{ n^2 - \frac{1}{3} L_z^2 \},
\end{eqnarray}
which is simply the interaction Hamiltonian for spinless $p$-orbital bosons~\cite{LiuWu2006pra}. Furthermore, we can reformulate the Eq.~\eqref{spin} in the bases of $p_\mu$, with $\mu =x,y$, and obtain that
\begin{equation}\label{intorigin}
H_{{I}} = U_{\text{int}} \sum_{\vec {i}} \{ n^2_{\vec {i}} +\frac{1}{2} \sum_{\mu \sigma} (n^2_{\vec i, \mu \sigma} + p^\dagger_{\vec i, \mu \sigma}p^\dagger_{\vec i,\mu \sigma} p_{\vec i,\bar \mu \sigma} p_{\vec i,\bar \mu \sigma}) \},
\end{equation}
where $U_{\text{int}} = \frac{U}{2}$, $n_{\vec i,\mu} = p^\dagger_{\vec i, \mu} p_{\vec i, \mu}$, and $n_{\vec i} = n_{\vec i, x} + n_{\vec i, y}$.

\section{Self consistent calculation and ground state}

Being of $C_4$ symmetry, the ground orbital state can be generically written as $e^{i\xi}(\phi_{p_x}+e^{i\varphi}\phi_{p_y})_\uparrow$ for spin-up component and $(\phi_{p_x}+e^{i\theta}\phi_{p_y})_\downarrow$ for spin-down component, where $\varphi$ and $\theta$ are the phases difference between $p_x$ and $p_y$ orbitals, with $-\pi/2 \leqslant \varphi,\theta \leqslant \pi/2$, and $\xi$ is the relative phase between spin-up and spin-down components. The phase factors can be determined by minimizing the energy of the total Hamiltonian including the single-particle SO coupled term and the interacting term
\begin{eqnarray}
H=H_0+H_{I}.
\end{eqnarray}
Before going to the self-consistent study, we analyze the effects of $H_0$ and $H_I$ separately. Having only the interacting Hamiltonian, the ground state corresponds to the solution with $\varphi =\theta= \pm \pi/2$ and arbitrary $\xi$. On the other hand, without the interacting term, the eigenstates of the SO coupled Hamiltonian $H_0$ correspond to the four band minimums and take the form
 \begin{equation}\label{singleeig}
\begin{pmatrix}
\phi_{p_x\uparrow} \\ \phi_{p_x\downarrow}
\end{pmatrix} = \begin{pmatrix} \sin \theta_{\rm sc} \\ \pm \cos\theta_{\rm sc} \end{pmatrix}, \quad \begin{pmatrix}
\phi_{p_y\uparrow} \\ \phi_{p_y\downarrow}
\end{pmatrix} = \begin{pmatrix} \sin \theta_{\rm sc} \\ \pm i\cos\theta_{\rm sc} \end{pmatrix}.
\end{equation}
where $\tan \theta_{\rm sc} = 2t_{so} \sin k_{\rm min}/(m_z + 2t_0 \cos k_{\rm min})$, with $k_{\rm min}$ being the momentum of one band-minima. From Eq.~\eqref{singleeig} we know that it is not possible for spin-up and spin-down component to both form local orbital configuration with $\theta = 0$ or $\pm \pi/2$. This reflects the competition between orbital degree of freedom and spin-orbit coupling.

For the self-consistent calculation, we define the mean-field order parameters by
\begin{eqnarray}
\Delta_\uparrow = \langle p^\dagger_{x(y)\uparrow}p_{x(y)\uparrow}\rangle,~\Delta_{\downarrow}=\langle p^\dagger_{x(y)\downarrow} p_{x(y)\downarrow} \rangle, ~ \Delta_{xy\uparrow} = \langle p^\dagger_{x\uparrow} p_{y\uparrow}  \rangle,~\Delta_{xy\downarrow} = \langle p^\dagger_{x\downarrow} p_{y\downarrow} \rangle.
\end{eqnarray}
It is trivial to know that $\Delta_{\uparrow(\downarrow)}$ is related to spin-polarization, while $\Delta_{xy\uparrow(\downarrow)}$ characterize the coupling between $p_x$ and $p_y$ orbitals, and are related to the aforementioned phase difference as $\Delta_{xy\uparrow(\downarrow)} = e^{i\varphi(\theta)}$. With the mean-field orders the interacting Hamiltonian can be linearized to
\begin{eqnarray}\label{mfh}
H_{\rm I}^{\rm MF}&=&U_{\text{int}}\sum_{\vec r} [3(n_{p_x\uparrow}\Delta_{\uparrow} +n_{p_x\downarrow}\Delta_\downarrow + n_{p_y\uparrow} \Delta_{\uparrow} +n_{p_y\downarrow} \Delta_{\downarrow})\nonumber\\
&+ &(\Delta_{xy\uparrow} p_{x\uparrow}^\dagger p_{y\uparrow} + \Delta_{xy\uparrow}^* p^\dagger_{y\uparrow} p_{x\uparrow}+\Delta_{xy \downarrow} p^\dagger_{x\downarrow}p_{y\downarrow} + \Delta_{xy\downarrow}^*p_{y\downarrow}^\dagger p_{x\downarrow})\nonumber \\
&+&2(\Delta_{xy\uparrow} p_{x\uparrow} p_{y\uparrow}^\dagger + \Delta_{xy\uparrow}^* p_{x\uparrow}^\dagger p_{y\uparrow} +\Delta_{xy\downarrow} p_{x\downarrow} p_{y\downarrow}^\dagger + \Delta_{xy\downarrow}^* p_{x\downarrow}^\dagger p_{y\downarrow})\nonumber\\
&+&4(n_{p_x\uparrow} \Delta_{\downarrow}  + \Delta_{\uparrow}n_{p_x\downarrow}   +  n_{p_y\uparrow}\Delta_\downarrow + \Delta_\uparrow n_{p_y\downarrow})].
\end{eqnarray}
The self-consistent solution can be obtained by the following iteration method. First, we solve the non-interacting ground state $\ket{G_0}$ for SO coupled Hamiltonian $H_0$. Then we calculate the mean-field order parameters based on $|G_0\rangle$. Furthermore, substitute the order parameters to mean-field Hamiltonian $H^{\text{MF}}=H_{0}+H_{\text{int}}^{\text{MF}}$ and solve the new ground state $\ket{G_{\text{MF}}}$, with which one can recalculate a new set of order parameters. Finally, we resubmit the new order parameters to the mean-field Hamiltonian $H^{\text{MF}}$, and repeat the calculation of $\ket{G_{\text{MF}}}$ and the mean-field orders until the solution converges.

With the self-consistently solved order parameters, we can also obtain the ground state wave function of the Bose-Einstein condensate (BEC). For this we define the bases $\hat b^\dagger_\uparrow =\frac{1}{\sqrt{2}} (p_x + e^{i\varphi}p_y^\dagger)_\uparrow$ and $\hat b^\dagger_\downarrow = \frac{1}{\sqrt{2}} (p_x + e^{i\theta}p_y)_\downarrow$ to rewrite the single particle Hamiltonian to be $H_s = \sum_{\bold k \sigma \sigma^\prime}\hat b^\dagger_{\sigma} \hat h_s(\bold k)^{\sigma\sigma'} b_{\sigma^\prime} $, where
\begin{equation}
\hat h_{s}(\bold{k}) = (m_z + t_0 \cos k_x + t_0 \cos k_y) \sigma_z - t_{so}(\sin k_x + \sin (\varphi - \theta)\sin k_y) \sigma_x - t_{so}\cos (\varphi - \theta) \sin k_y \sigma_y.
\end{equation}
On the other hand,
the original interacting Hamiltonian Eq.~\eqref{intorigin} can be projected onto orbital bases $\hat b_{\sigma}^\dagger$. We find that
\begin{equation}
H_I^\prime = U_{\text{int}} \sum_{\vec i} (n^2_{\vec i} + \frac{1}{2}n^2_{\vec i, \uparrow} \cos^2 \varphi + \frac{1}{2}n^2_{\vec i, \downarrow} \cos^2 \theta),
\end{equation}
where $n_{\vec i, \uparrow} = b^\dagger_{\vec i \uparrow} b_{\vec i\uparrow}$, $n_{\vec i, \downarrow} = b^\dagger_{\vec i \downarrow} b_{\vec i \downarrow}$ and $n_{\vec i} = n_{\vec i,\uparrow} + n_{\vec i, \downarrow}$. Under the mean-field approximation, the interacting Hamiltonian can be recast into
\begin{equation}
H_{I}^{\text{MF}} = U_{\text{int}} \bigr[2n (\Delta_\uparrow + \Delta_\downarrow) + n_\uparrow\Delta_\uparrow \cos^2 \varphi + n_\downarrow \Delta_\downarrow \cos^2 \theta\bigr],
\end{equation}
where we have already perform the Fourier transformation for a certain band minima. The above interaction further modifies the effective Zeeman field $m_z$ to $\tilde m_z = m_z + \delta m_z$ with
\begin{equation}
	\delta m_z = \frac{U_{\text{int}}}{2}(\Delta_\uparrow \cos^2 \varphi - \Delta_\downarrow \cos^2 \theta)\sigma_z.
\end{equation}
With the above results we finally reach the total mean-field Hamiltonian, which is given by $H_{\text{MF}} =H_s+H_I^{\rm MF} = \sum_{\bold k \sigma \sigma^\prime}\hat b^\dagger_{\sigma} \hat h_{\text{MF}}^{\sigma\sigma'}(\bold k) b_{\sigma^\prime} $, and
\begin{equation}
\hat h_{\text{MF}} (\bold{k}) = \vec \lambda \cdot{} \vec \sigma,
\end{equation}
where
\begin{eqnarray}
\lambda_x &=& -t_{so}(\sin k_x + \sin (\varphi - \theta) \sin k_y), \\
\lambda_y &=& -t_{so} \cos (\varphi - \theta) \sin k_y, \\
\lambda_z &=& \tilde m_z + t_0 \cos k_x + t_0 \cos k_y.
\end{eqnarray}
Thus the eigenvalues for $\hat h_0(\bold{k})$ are
\begin{equation}
\epsilon_{\pm} = \pm \lambda (\bold{k}), \quad \lambda = \sqrt{\lambda_x^2 + \lambda_y^2 + \lambda_z^2}.
\end{equation}
The corresponding two eigenstates of Hamiltonian are given by
\begin{equation}
\ket{\mu_{k,+}} = \begin{pmatrix} \cos \frac{\alpha(\bold{k})}{2} e^{i\beta(\bold{k})} \\ \sin \frac{\alpha(\bold{k})}{2}\end{pmatrix} \quad \ket{\mu_{k,-}} = \begin{pmatrix} \sin \frac{\alpha(\bold{k})}{2} e^{i\beta(\bold{k})} \\ -\cos \frac{\alpha(\bold{k})}{2}\end{pmatrix}
\end{equation}
where $\tan \alpha=\lambda_z / \sqrt{\lambda_x^2 + \lambda_y^2}$ and $\tan \beta = \lambda_y / \lambda_x$. When $t_{so}$ much more larger than $t_0$, the system has four band minima which are very close to $(\pm \pi/2, \pm \pi/2)$. The single-particle ground state is $\ket{\mu_{k,-}}$ at one of the four band minima, and the BEC is obtained by condensing the bosons at the band minimum, given by
\begin{eqnarray}\label{condensate1}
|\Phi_{\rm BEC}\rangle=\sin\alpha e^{i\beta}|\Phi_{p_x+e^{i\varphi}p_y},\uparrow\rangle-\cos\alpha|\Phi_{p_x+e^{i\theta}p_y},\downarrow\rangle.
\end{eqnarray}
It can be verified straightforwardly that the parameter $\xi$ defined previously is given by $\xi=\beta$.

\section{Excitation band structure}

Now we discuss the calculation of phonon excitations. The chemical potential $\mu$ in Bogoliubov Hamiltonian can be determined numerically through imaginary time evolution \cite{BaderJCP2013}. For a Hermitian operator, like Hamiltonian, we can always have positive definite eigenvalues by properly choosing the minimum of the potential. We consider that
\begin{equation}
0 \leq E_0 \leq E_1 \leq E_2 .....
\end{equation}
For the time dependent Schrodinger Equation:
\begin{equation}
i \frac{\partial}{\partial t} \psi(\vec r,t) = H \psi(\vec r,t), \quad \psi(\vec r,0)=\psi_{0}(\vec r)
\end{equation}
After a Wick rotation of the time coordinate $t = -i\tau $, we can then derive:
\begin{equation}
- \frac{\partial}{\partial \tau} \psi(\vec r,t) = H \psi(\vec r,\tau), \quad \psi(\vec r,0)=\psi_{0}(\vec r)
\end{equation}
The solution is:
\begin{equation}
\psi(\vec r,\tau) = e^{-\tau H} \psi(\vec r,0)
\end{equation}
On the other hand, the initial state can be expanded by eigenstates as: \begin{equation}
\psi_0(\vec r) = \sum_i c_i \phi_i(\vec r), \quad c_i = \langle \phi_i(\vec r)| \psi(\vec r,0)\rangle
\end{equation}
Thus we have
\begin{equation}
\psi(\vec r,\tau) = e^{-\tau H} \psi(\vec r,0)=\sum_i e^{- \tau E_i} c_i \phi_i(\vec r)
\end{equation}
After a sufficient long time, $\psi(\vec r,\tau)$ asymptotically approaches to the ground state as $e^{-\tau E_0}c_0 \phi_0$~\cite{BaderJCP2013}. The chemical potential is determined by the ground state energy of the BEC.

The interacting part of our Hamiltonian can modify the Zeeman term $m_z$ to an effective, and thus change the band gap. The band gap of excitation spectrum can be estimated in the following way.
The diagonal terms in $\Gamma_1$ in main text can be written as
\begin{equation}
\begin{aligned}
\text{diag} (\Gamma_1) &= \begin{pmatrix}
2\tilde g_{\uparrow\uparrow} |\phi_{\uparrow}|^2+\tilde g_{\uparrow\downarrow}|\phi_\downarrow|^2 & \\
 & \tilde g_{\uparrow\downarrow}|\phi_{\uparrow}|^2+2\tilde g_{\downarrow\downarrow}|\phi_\downarrow|^2
\end{pmatrix} \\
&= \frac{1}{2}(2\tilde g_{\uparrow\uparrow} |\phi_{\uparrow}|^2+\tilde g_{\uparrow\downarrow}|\phi_\downarrow|^2-\tilde g_{\uparrow\downarrow}|\phi_{\uparrow}|^2-2\tilde g_{\downarrow\downarrow}|\phi_\downarrow|^2)\sigma_z + \frac{1}{2}(2\tilde g_{\uparrow\uparrow} |\phi_{\uparrow}|^2+\tilde g_{\uparrow\downarrow}|\phi_\downarrow|^2+\tilde g_{\uparrow\downarrow}|\phi_{\uparrow}|^2+2\tilde g_{\downarrow\downarrow}|\phi_\downarrow|^2) I
\end{aligned}
\end{equation}
Note that $g_{\uparrow\downarrow} = g$, $g_{\uparrow\uparrow} = g(1 + \frac{1}{2} \cos^2 \varphi)$, and $g_{\downarrow\downarrow} = g(1 + \frac{1}{2} \cos^2 \theta)$. The term regarding $\sigma_z$ can be written as
\begin{equation}
M_f(\varphi) \sigma_z = \frac{g}{2}(|\phi_{\uparrow}|^2-|\phi_\downarrow|^2 + \cos^2 \varphi |\phi_\uparrow|^2-\cos^2 \theta |\phi_\downarrow|^2)\sigma_z
\end{equation}
This term can modify $m_z$ effectively and thus can influence the band gap for topological phonons. Two cases can be obtained. First, if $M_f(\varphi,\theta) = (|\phi_{\uparrow}|^2-|\phi_\downarrow|^2 + \cos^2 \varphi |\phi_\uparrow|^2-\cos^2 \theta |\phi_\downarrow|^2) > 0$, then this term can increase the $E_{\text{gap}}$ effectively. Secondly, if $M_f(\varphi,\theta) = (|\phi_{\uparrow}|^2-|\phi_\downarrow|^2 + \cos^2 \varphi |\phi_\uparrow|^2-\cos^2 \theta |\phi_\downarrow|^2)< 0$, then this term can decrease the $E_{\text{gap}}$ effectively.
We can then estimated the band gap as
\begin{equation}
E_{\text{Estimated Gap}} = 2\bigr[m_z+M_f(\varphi)\bigr] =2m_z+g(|\phi_{\uparrow}|^2-|\phi_\downarrow|^2 + \cos^2 \varphi |\phi_\uparrow|^2-\cos^2 \theta |\phi_\downarrow|^2).
\end{equation}
Note that before the phase transition ($U_{\rm int}<U_{c1}$), the Dirac cone is protected and the phonons keep to be gapless.


\end{document}